\documentclass[sigconf,natbib=true]{acmart}
\usepackage{enumitem}
\usepackage{multirow}
\usepackage{booktabs}
\usepackage{subcaption} % support subfigure env
\usepackage{hyperref}
\usepackage{colortbl}
\usepackage{graphicx}   % 插入图片
\usepackage{makecell}   % 支持单元格多行
\usepackage{xcolor}

\AtBeginDocument{%
  }

\copyrightyear{2026}
\acmYear{2026}
\setcopyright{cc}
\setcctype{by}
\acmConference[SIGIR '26] {Proceedings of the 49th International ACM SIGIR Conference on Research and Development in Information Retrieval}{July 20--24, 2026}{Melbourne, VIC, Australia.}
\acmBooktitle{Proceedings of the 49th International ACM SIGIR Conference on Research and Development in Information Retrieval (SIGIR '26), July 20--24, 2026, Melbourne, VIC, Australia}
\acmISBN{979-8-4007-2599-9/2026/07}
\acmDOI{10.1145/3805712.3809892}
\begin{document}

\title{RoTE: Coarse-to-Fine Multi-Level Rotary Time Embedding for Sequential Recommendation}


\author{Haolin Zhang}
\authornote{Both authors contributed equally to this research.}
\affiliation{%
  \institution{Shenzhen International Graduate School, Tsinghua University}
  \city{Shenzhen}
  \country{China}
}
\email{zhanghaolin24@mails.tsinghua.edu.cn}

\author{Longtao Xiao}
\authornotemark[1]
\affiliation{%
  \institution{Huazhong University of Science and Technology}
  \city{Wuhan}
  \country{China}
}
\email{xiaolongtao@hust.edu.cn}

\author{Guohao Cai}
\affiliation{%
  \institution{Huawei Noah’s Ark Lab}
  \city{Shenzhen}
  \country{China}
}
\email{caiguohao1@huawei.com}

\author{Ruixuan	Li}
\affiliation{%
  \institution{Huazhong University of Science and Technology}
  \city{Wuhan}
  \country{China}
}
\email{rxli@hust.edu.cn}

\author{Xiu Li}
\affiliation{%
  \institution{Shenzhen International Graduate School, Tsinghua University}
  \city{Shenzhen}
  \country{China}
}
\email{li.xiu@sz.tsinghua.edu.cn}
\authornote{Corresponding author}

\renewcommand{\shortauthors}{Zhang et al.}

\begin{abstract}
Sequential recommendation models have been widely adopted for modeling user behavior. Existing approaches typically construct user interaction sequences by sorting items according to timestamps and then model user preferences from historical behaviors. While effective, such a process only considers the order of temporal information but overlooks the actual time spans between interactions, resulting in a coarse representation of users’ temporal dynamics and limiting the model’s ability to capture long-term and short-term interest evolution. To address this limitation, we propose \textbf{RoTE}, a novel multi-level temporal embedding module that explicitly models time span information in sequential recommendation. RoTE decomposes each interaction timestamp into multiple temporal granularities, ranging from coarse to fine, and incorporates the resulting temporal representations into item embeddings. This design enables models to capture heterogeneous temporal patterns and better perceive temporal distances among user interactions during sequence modeling. RoTE is a lightweight, plug-and-play module that can be seamlessly integrated into existing Transformer-based sequential recommendation models without modifying their backbone architectures. We apply RoTE to several representative models and conduct extensive experiments on three public benchmarks. Experimental results demonstrate that RoTE consistently enhances the corresponding backbone models, achieving up to a 20.11\% improvement in NDCG@5, which confirms the effectiveness and generality of the proposed approach. Our code is available at https://github.com/XiaoLongtaoo/RoTE.

\end{abstract}





\maketitle

\section{Introduction}

Sequential recommendation \cite{seqrec_study, pan2026survey, wei2025sequential} is a fundamental problem in recommender systems, which aims to predict the next item a user will interact with based on their historical behavior sequences. Sequential recommender models \cite{fpmc, covington2016deep, shah2025dsrs, he2016fusing} capture dynamic user preferences by modeling interaction sequences, constructed by chronologically ordering previously interacted items. Existing methods can be broadly categorized into traditional (discriminative) approaches \cite{elecrec} and generative approaches \cite{xiao2025unger}. Traditional sequential recommendation \cite{gru4rec, caser, sasrec, bert4rec} formulate the recommendation task as a candidate matching task, in which the model learns user preferences from historical interaction sequences and derives latent user interest representations, which are subsequently utilized to calculate the relevance of candidate items, thus guiding the ranking and selection of items. Generative sequential recommendation \cite{www25-gen-rec-tutorial, hou2025survey} treats recommendation as a sequence generation problem, modeling user behavior in an autoregressive or sequence-to-sequence manner and often leveraging large language models \cite{p5, gptrec} or sequence generation frameworks \cite{tiger, hstu} to achieve more flexible and unified representation learning.

% Despite the strong empirical performance of traditional and generative sequential recommendation models, a key limitation persists in how they handle temporal information. Most existing traditional and generative methods typically sort interaction sequences by timestamp and then rely on positional embeddings or relative order as a coarse proxy for temporal information to capture sequence structure, where such positional encoding can be viewed as a highly simplified form of temporal modeling that implicitly assumes uniformly spaced interactions. This assumption ignores the true time intervals between interactions, results in an oversimplified representation of temporal dynamics and inhibits the model’s ability to capture nuanced temporal patterns such as drift in preferences over irregular time gaps.

Despite the strong empirical performance of traditional and generative sequential recommendation models, a key limitation \cite{nasir2023survey} persists in how they handle temporal information. Most existing traditional and generative methods typically sort interaction sequences by timestamp and then rely on positional embeddings to capture temporal information. Such positional encoding \cite{lopez2024positional} can be viewed as the most simplified form of temporal modeling, as it primarily preserves relative order while treating time as uniformly spaced and ignoring heterogeneous time intervals between interactions. As a result, rich temporal signals at different time scales are either entangled or discarded, leading to an oversimplified representation of temporal dynamics and limiting the model’s ability to capture nuanced patterns such as multi-scale preference evolution and drift over irregular time gaps.

In real-world scenarios, interaction time is commonly recorded in the form of Unix timestamps \cite{hauser2018unix}, providing a precise and continuous measure of when user actions occur. However, such representations do not align well with how users perceive and organize time in daily life. Instead of reasoning in terms of absolute numerical values, users naturally interpret time in a hierarchical and semantic manner \cite{poppel1997hierarchical}, where long-term interests evolve across years, medium-term behaviors are shaped by seasonal or monthly patterns, and short-term intents are driven by daily interactions. This mismatch between the raw timestamp representation and human time perception introduces a bias in temporal modeling, as heterogeneous temporal signals at different scales are compressed into a single continuous value. 

To address these challenges, we propose \textbf{Ro}tary \textbf{T}ime \textbf{E}mbedding (RoTE), a lightweight and plug-and-play multi-level temporal modeling module for sequential recommendation. RoTE explicitly captures coarse-to-fine temporal information and encodes them into item embeddings in the self-attention module. Specifically, RoTE decomposes interaction timestamps into year, month, and day, and integrates them into item representations through a unified rotary embedding mechanism. For each temporal level, RoTE injects time-aware signals directly into the item embedding space, enabling temporal information to be tightly coupled with sequence modeling rather than treated as auxiliary features. The resulting time-enhanced representations are then aggregated to form the final item embeddings used by the model. By seamlessly integrating multi-level temporal signals into the self-attention process without modifying backbone architectures, RoTE allows existing Transformer-based sequential recommendation models to better perceive heterogeneous temporal distances among user interactions without introducing significant computational overhead.

The key contributions of our work are as follows:

\begin{itemize}[left=0pt]

    \item We propose RoTE, a novel plug-and-play temporal embedding module that explicitly models time-span information for sequential recommendation.

    \item We introduce a coarse-to-fine temporal encoding strategy that enables Transformer-based sequential recommendation models to capture multi-level temporal patterns.

    \item We demonstrate that RoTE consistently improves recommendation performance through extensive experiments on multiple benchmark datasets and representative models, showcasing its advantages in performance and generality.
    
\end{itemize}

\section{Methodology}

\subsection{Problem Definition}

Let $\mathcal{U}$ and $\mathcal{I}$ denote the sets of users and items, respectively.
Each user $u \in \mathcal{U}$ is associated with an interaction sequence
\begin{equation}
\mathcal{S}_u = \{(i_0, t_0), (i_1, t_1), \dots, (i_{L_u-1}, t_{L_u-1})\},
\end{equation}
where $i_k \in \mathcal{I}$ denotes the $k$-th interacted item and $t_k$ is the corresponding interaction timestamp, satisfying $t_0 < t_1 < \dots < t_{L_u-1}$.
The sequence length $L_u$ may vary across users, and the time intervals between consecutive interactions are generally irregular.

Given a user’s historical interaction sequence $\mathcal{S}_u$, the goal of sequential recommendation is to model the dependency between past interactions and future behaviors, and predict the next item $i_{L_u}$ conditioned on the observed sequence.
Formally, the task can be defined as learning a conditional distribution
\begin{equation}
p(i_{L_u} \mid \mathcal{S}_u),
\end{equation}
which can be instantiated either as a ranking-based scoring function or as a generative model that directly produces the next item token.

\begin{figure}
    \centering
    \includegraphics[width=\linewidth]{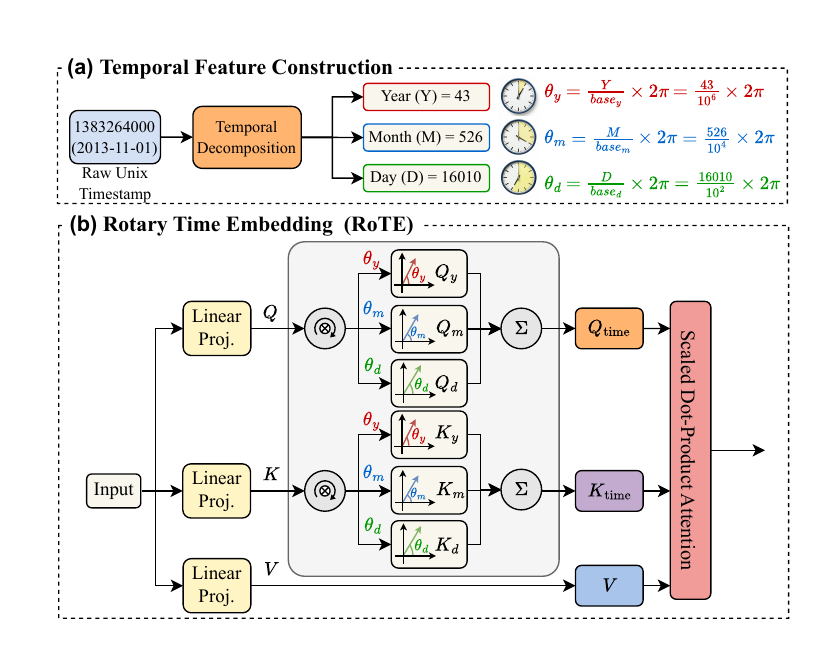}
    \caption{Illustration of the proposed RoTE module.}
    \label{fig:RoTE}
\end{figure}

In this work, we focus on enhancing sequential recommendation models by more effectively modeling the temporal information embedded in interaction timestamps, as shown in Figure \ref{fig:RoTE}.
Specifically, we aim to capture heterogeneous time intervals and multi-scale temporal dynamics in a unified manner that is compatible with Transformer-based generative and discriminative sequential recommendation architectures.

\subsection{Temporal Feature Construction}

In real-world scenarios, interaction time is typically recorded as a Unix timestamp. To obtain a structured temporal representation, we decompose each timestamp $t_k$ into a hierarchical triplet
\begin{equation}
t_k \rightarrow (y_k, m_k, d_k),
\end{equation}
where $y_k$, $m_k$, and $d_k$ denote the cumulative number of years, months, and days elapsed since January 1, 1970, respectively.
For example, January 1, 1970 corresponds to $(0, 0, 0)$, while January 1, 1971 corresponds to $(1, 12, 365)$. 
This ordinal representation preserves temporal ordering and enables modeling temporal dynamics across multiple granularities, providing a foundation for for temporal encoding.

\subsection{Rotary Time Embedding (RoTE)}

Based on the constructed multi-level temporal features, we propose \textbf{Ro}tary \textbf{T}ime \textbf{E}mbedding (RoTE), a lightweight temporal modeling module that injects time information directly into item representations within the self-attention mechanism. RoTE integrates time-aware signals into the item embedding space by applying rotary-style transformations \cite{rope} to the query and key representations, allowing the attention mechanism to perceive heterogeneous temporal distances among interactions.

Given an input sequence of item embeddings, we first project them into query, key, and value representations following standard multi-head self-attention.
For each interaction, its timestamp is represented by a hierarchical  triplet $(y_k, m_k, d_k)$.
For each temporal level, RoTE computes rotation angles using a fixed inverse frequency spectrum determined by a level-specific base scalar.
Specifically, for a head dimension $d$, the inverse frequency vector $\boldsymbol{\omega}^{(l)} \in \mathbb{R}^{d/2}$ in a given temporal level $l \in \{y, m, d\}$ is defined as
\begin{equation}
\boldsymbol{\omega}^{(l)}_i = \left(\text{base}^{(l)}\right)^{- \frac{2i}{d}}, \quad i = 0, \dots, \frac{d}{2}-1,
\end{equation}
where $\text{base}^{(l)}$ is a predefined scalar controlling the frequency range for temporal level $l$.
Given the decomposed temporal value $y_k, m_k, d_k$, the corresponding rotation angle is computed as
\begin{equation}
\boldsymbol{\theta}_k^{(y)} = y_k \cdot \boldsymbol{\omega}^{(y)}, \quad
\boldsymbol{\theta}_k^{(m)} = m_k \cdot \boldsymbol{\omega}^{(m)}, \quad
\boldsymbol{\theta}_k^{(d)} = d_k \cdot \boldsymbol{\omega}^{(d)}.
\end{equation}

The rotation angles are then used to inject temporal information into the query and key representations via rotary transformations.
Specifically, the query and key vectors are rotated in a pairwise manner over even-odd dimensions.
For each dimension pair $(2i, 2i+1)$, the rotation can be interpreted as applying a 2D rotation matrix parameterized by $\theta_{k,i}^{(l)}$:
\begin{equation}
\begin{pmatrix}
q_{k,2i}^{(l)} \\
q_{k,2i+1}^{(l)}
\end{pmatrix}
=
\begin{pmatrix}
\cos \theta_{k,i}^{(l)} & -\sin \theta_{k,i}^{(l)} \\
\sin \theta_{k,i}^{(l)} & \cos \theta_{k,i}^{(l)}
\end{pmatrix}
\begin{pmatrix}
q_{k,2i} \\
q_{k,2i+1}
\end{pmatrix},
\end{equation}
and the same transformation is applied to the key vector $\mathbf{k}_k$.

For notational simplicity, the rotary transformation can be written in a vectorized form as
\begin{equation}
\begin{aligned}
\mathbf{q}_k^{(l)} &= \mathbf{q}_k \odot \cos \boldsymbol{\theta}_k^{(l)} 
+ \mathrm{rotate}(\mathbf{q}_k) \odot \sin \boldsymbol{\theta}_k^{(l)}, \\
\mathbf{k}_k^{(l)} &= \mathbf{k}_k \odot \cos \boldsymbol{\theta}_k^{(l)} 
+ \mathrm{rotate}(\mathbf{k}_k) \odot \sin \boldsymbol{\theta}_k^{(l)},
\end{aligned}
\end{equation}
where $\mathrm{rotate}(\cdot)$ denotes a fixed permutation operator that rearranges the input vector by swapping each even-odd dimension pair.
Specifically, for an input vector $\mathbf{x} \in \mathbb{R}^{d}$, the operator is defined as
\begin{equation}
\mathrm{rotate}(\mathbf{x}) = (-x_1, x_0, -x_3, x_2, \dots, -x_{d-1}, x_{d-2}),
\end{equation}
which enables a vectorized implementation of the 2D rotation in each $(2i, 2i+1)$ subspace.This operation preserves vector norms while encoding temporal signals directly into the angular relationships between queries and keys, thereby enabling attention scores to be sensitive to relative temporal differences.

To capture temporal dynamics at multiple time scales, RoTE applies rotary transformations independently for the year-, month-, and day-level temporal components, producing three time-aware representations for each query and key.
These representations are then aggregated via a weighted fusion strategy:
\begin{equation}
\begin{aligned}
\mathbf{q}_k^{\text{time}} &=
\alpha_y \mathbf{q}_k^{(y)} +
\alpha_m \mathbf{q}_k^{(m)} +
\alpha_d \mathbf{q}_k^{(d)}, \\
\mathbf{k}_k^{\text{time}} &=
\alpha_y \mathbf{k}_k^{(y)} +
\alpha_m \mathbf{k}_k^{(m)} +
\alpha_d \mathbf{k}_k^{(d)},
\end{aligned}
\end{equation}
where $\alpha_y$, $\alpha_m$, and $\alpha_d$ are scalar weights that control the relative contributions of different temporal levels.
In our implementation, larger weights are assigned to coarse-grained temporal components to emphasize long-term preference evolution, while finer-grained components focus on short-term dynamics.

The resulting time-enhanced query and key representations are then used to compute attention scores using the standard scaled dot-product attention.
Importantly, RoTE does not modify the value representations, attention formulation, or training objective, and can be seamlessly integrated into existing Transformer-based sequential recommendation models.
This design makes RoTE a plug-and-play temporal modeling module that enhances temporal awareness without incurring significant computational overhead.

% \subsection{Integration with Sequential Recommendation Models}

% RoTE can be seamlessly integrated into both traditional and generative Transformer-based sequential recommendation models. 
% In practice, each item in the input sequence is augmented with its decomposed temporal features $(y_k, m_k, d_k)$, which serve as additional signals for temporal modeling. 
% Within the self-attention layers, the standard positional encoding is replaced by RoTE-based time-aware rotary transformations applied to the query and key representations. 
% By this way, RoTE can be applied broadly across different sequential recommendation frameworks, including both traditional paradigm and generative paradigm.

\section{Experiments}

\subsection{Experimental Setting}

\subsubsection{Dataset.}

We evaluate RoTE on three public benchmarks from the Amazon Reviews dataset \cite{amazon}: \textit{Sports and Outdoors}, \textit{Beauty}, and \textit{Toys and Games}. 
Following common practice in recent sequential recommendation research \cite{tiger}, we adopt the 5-core versions of these datasets, in which each user and each item has at least five interactions, thereby improves data quality and reduces noise in sparse interaction histories. 
Each dataset is chronologically sorted by timestamp, and sequences are truncated or padded to a fixed maximum length of 50 for both training and evaluation. Table \ref{tab:statistics} presents the detailed statistics of the three datasets.

\begin{table}[t]
    \centering
    \caption{Statistics of the Datasets.}
    \label{tab:statistics}
    \resizebox{\linewidth}{!}{
    \begin{tabular}{ccccc}
        \toprule
        Dataset & \#Users & \#Items & \#Interactions & \#Density \\
        \midrule
        Beauty & 22,363 & 12,101 & 198,360 & 0.00073 \\
        Sports and Outdoors & 35,598 & 18,357 & 296,175 & 0.00045 \\
        Toys and Games & 19,412 & 11,924 & 167,526 & 0.00073 \\
        \bottomrule
    \end{tabular}
    }
\end{table}

\subsubsection{Baselines.}

We compare RoTE with the following baseline models, covering both traditional and generative paradigms. 
Traditional baselines \cite{gru4rec, caser, bert4rec, sasrec} capture sequential dependencies in user behavior via next-item prediction, while generative baselines \cite{vqrec, tiger, hstu, rpg} adopt autoregressive generation frameworks to model user interaction sequences. 
To evaluate the effectiveness of RoTE as a plug-and-play temporal modeling module, we integrate it into selected representative Transformer-based backbones from each paradigm, while other baselines are reported in their original forms. 

\begin{table*}[t]
\caption{Overall performance comparison of state-of-the-art sequential recommendation models and their RoTE-enhanced counterparts on three benchmark datasets. RoTE is a plug-and-play module that improves model performance without modifying the backbone architectures. Performance improvements of RoTE-based models over their corresponding backbones are statistically significant according to a paired t-test with $p < 0.05$.}
\centering
\setlength{\tabcolsep}{4pt}
\begin{tabular}{
  l|l % 这条 | 会应用到表格的每一行
  c c c c
  c c c c
  c c c c
}
\toprule
% --- 表头 ---
\multicolumn{1}{l}{} & \multirow{2.5}{*}{\textbf{Methods}} & \multicolumn{4}{c}{\textbf{Sports and Outdoors}} & \multicolumn{4}{c}{\textbf{Beauty}}  & \multicolumn{4}{c}{\textbf{Toys and Games}} \\
\cmidrule(lr){3-6}\cmidrule(lr){7-10}\cmidrule(lr){11-14}
\multicolumn{1}{l}{} & & R@5 & N@5 & R@10 & N@10 & R@5 & N@5 & R@10 & N@10 & R@5 & N@5 & R@10 & N@10 \\
\midrule
% --- Traditional 组 ---
\multirow{6}{*}{\rotatebox{90}{\textbf{Traditional}}}
& GRU4Rec     & 0.0129 & 0.0086 & 0.0204 & 0.0110 & 0.0164 & 0.0099 & 0.0283 & 0.0137 & 0.0097 & 0.0059 & 0.0176 & 0.0084 \\
& Caser       & 0.0116 & 0.0072 & 0.0194 & 0.0097 & 0.0205 & 0.0131 & 0.0347 & 0.0176 & 0.0166 & 0.0107 & 0.0270 & 0.0141 \\
& BERT4Rec    & 0.0115 & 0.0075 & 0.0191 & 0.0099 & 0.0203 & 0.0124 & 0.0347 & 0.0170 & 0.0116 & 0.0071 & 0.0203 & 0.0099 \\
& SASRec      & 0.0228 & 0.0146 & 0.0347 & 0.0189 & 0.0375 & 0.0238 & 0.0604 & 0.0311 & 0.0466 & 0.0310 & 0.0688 & 0.0375 \\
& \textbf{RoTE-SASRec} & \textbf{0.0233} & \textbf{0.0158} & \textbf{0.0372} & \textbf{0.0202} & \textbf{0.0400} & \textbf{0.0260} & \textbf{0.0638} & \textbf{0.0336} & \textbf{0.0512} & \textbf{0.0345} & \textbf{0.0747} & \textbf{0.0420} \\
& \textit{Improv.} & \textcolor{green!50!black}{\textit{+2.19\%}} & \textcolor{green!50!black}{\textit{+8.22\%}} & \textcolor{green!50!black}{\textit{+7.20\%}} & \textcolor{green!50!black}{\textit{+6.88\%}} & \textcolor{green!50!black}{\textit{+6.67\%}} & \textcolor{green!50!black}{\textit{+9.24\%}} & \textcolor{green!50!black}{\textit{+5.63\%}} & \textcolor{green!50!black}{\textit{+8.04\%}} & \textcolor{green!50!black}{\textit{+9.87\%}} & \textcolor{green!50!black}{\textit{+11.29\%}} & \textcolor{green!50!black}{\textit{+8.58\%}} & \textcolor{green!50!black}{\textit{+12.00\%}} \\
\midrule
% --- Generative 组 ---
\multirow{6}{*}{\rotatebox{90}{\textbf{Generative}}}
& VQ-Rec      & 0.0208 & 0.0137 & 0.0300 & 0.0173 & 0.0457 & 0.0317 & 0.0664 & 0.0383 & 0.0373 & 0.0256 & 0.0484 & 0.0323 \\
& TIGER       & 0.0264 & 0.0181 & 0.0400 & 0.0225 & 0.0454 & 0.0321 & 0.0648 & 0.0384 & 0.0521 & 0.0371 & 0.0712 & 0.0432 \\
& HSTU        & 0.0258 & 0.0165 & 0.0414 & 0.0215 & 0.0469 & 0.0314 & 0.0704 & 0.0389 & 0.0433 & 0.0281 & 0.0669 & 0.0357 \\
& RPG         & 0.0294 & 0.0201 & 0.0419 & 0.0241 & 0.0504 & 0.0349 & 0.0730 & 0.0422 & 0.0531 & 0.0373 & 0.0759 & 0.0446 \\
& \textbf{RoTE-RPG}    & \textbf{0.0317} & \textbf{0.0222} & \textbf{0.0456} & \textbf{0.0266} & \textbf{0.0521} & \textbf{0.0370} & \textbf{0.0742} & \textbf{0.0438} & \textbf{0.0624} & \textbf{0.0448} & \textbf{0.0852} & \textbf{0.0521} \\
& \textit{Improv.} & \textcolor{green!50!black}{\textit{+7.82\%}} & \textcolor{green!50!black}{\textit{+10.45\%}} & \textcolor{green!50!black}{\textit{+8.83\%}} & \textcolor{green!50!black}{\textit{+10.37\%}} & \textcolor{green!50!black}{\textit{+3.37\%}} & \textcolor{green!50!black}{\textit{+6.02\%}} & \textcolor{green!50!black}{\textit{+1.64\%}} & \textcolor{green!50!black}{\textit{+3.79\%}} & \textcolor{green!50!black}{\textit{+17.51\%}} & \textcolor{green!50!black}{\textit{+20.11\%}} & \textcolor{green!50!black}{\textit{+12.26\%}} & \textcolor{green!50!black}{\textit{+16.82\%}} \\
\bottomrule
\end{tabular}
\label{tab:main_results}
\end{table*}

\subsubsection{Implementation Details.}

We implement RoTE module using the PyTorch framework. 
For generative models, we adopt RPG \cite{rpg} as a representative backbone and build our implementation upon its official codebase\footnote{https://github.com/facebookresearch/RPG\_KDD2025}. 
For traditional models, we select SASRec \cite{sasrec, sasrec_revis} as a representative and use the publicly available implementation\footnote{https://github.com/pmixer/SASRec.pytorch}. Based on these backbones, we replace the original position encoding module with RoTE while keeping the remaining architecture unchanged. 
For the year-, month-, and day-level temporal components, the base frequencies are set to $\text{base}^{(y)} = 10^6$, $\text{base}^{(m)} = 10^4$, and $\text{base}^{(d)} = 10^2$, respectively, and the corresponding fusion weights are set to $\alpha_y = 1.5$, $\alpha_m = 1.0$, and $\alpha_d = 0.5$. 
Except for the integration of RoTE, all other architectural configurations follow the standard setups reported in the original implementations.

\subsubsection{Evaluation Metrics.}

We evaluate model performance using Recall@K and NDCG@K, where $K \in \{5, 10\}$. 
Following the standard leave-one-out evaluation strategy, we reserve the last interaction of each user as the test item, the penultimate item for validation, and all remaining interactions for training. Results are averaged over five random seeds.

\subsection{Overall Performance}

Table \ref{tab:main_results} reports the overall performance of representative sequential recommendation models and their RoTE-enhanced variants on three Amazon benchmarks.
As shown in the results, incorporating RoTE consistently improves the performance of both traditional and generative models across all datasets and evaluation metrics.
Specifically, RoTE brings stable gains over strong baselines, including SASRec and RPG, achieving noticeable improvements in performance. Notably, on the \textit{Toys and Games} benchmark, RoTE improves Recall@5 and NDCG@5 by 17.51\% and 20.11\% over the RPG baseline, respectively.
These improvements indicate that RoTE effectively captures temporal information and can be seamlessly integrated into diverse sequential modeling paradigms.
Overall, the results demonstrate the effectiveness and broad applicability of RoTE as a plug-and-play module for sequential recommendation.

\subsection{Ablation Study}

\begin{table}[t]
\centering
\caption{Ablation study of different levels of temporal signals on \textit{Toys and Games} dataset.}
\label{tab:ablation}
\begin{tabular}{lccccc}
\toprule
\multirow{2}{*}{\textbf{Methods}} 
& \multicolumn{2}{c}{\textbf{SASRec}} 
& \multicolumn{2}{c}{\textbf{RPG}} \\
\cmidrule(lr){2-3} \cmidrule(lr){4-5}
& \textbf{R@10} & \textbf{N@10} 
& \textbf{R@10} & \textbf{N@10} \\
\midrule
Backbone      & 0.0688 & 0.0375 & 0.0759 & 0.0446 \\
+ Pure Timestamp          & 0.0713 & 0.0403 & 0.0838 & 0.0498 \\
+ Y                       & 0.0701 & 0.0387 & 0.0815 & 0.0467 \\
+ Y + M                     & 0.0725 & 0.0409 & 0.0846 & 0.0505 \\
\textbf{+ Y + M + D (RoTE)}      & \textbf{0.0747} & \textbf{0.0420} & \textbf{0.0852} & \textbf{0.0521} \\
\bottomrule
\end{tabular}
\end{table}

Table \ref{tab:ablation} reports ablation results with different temporal encoding strategies on \textit{Toys and Games}. 
\textbf{Backbone} denotes the original models using positional embeddings, which only preserve relative order and can be viewed as a degenerate form of temporal modeling. 
\textbf{Pure Timestamp} extends this formulation by directly applying rotary embedding to absolute Unix timestamps.
The \textbf{Y}, \textbf{Y+M}, and \textbf{Y+M+D} variants progressively introduce year-, month-, and day-level calendar components into RoTE. 
Results show that while absolute timestamps provide measurable improvements over positional encoding, performance consistently improves as finer-grained and semantically structured temporal components are incorporated. 
\textbf{Y+M+D} achieves the best results on both SASRec and RPG, validating the effectiveness of coarse-to-fine multi-level temporal modeling.

\subsection{Efficiency Analysis}

\begin{table}[t]
\centering
\caption{Efficiency analysis of RoTE on different backbones.}
\label{tab:efficiency}
\begin{tabular}{lccc}
\toprule
\textbf{Methods} & \textbf{\#Params} & \textbf{FLOPs} & \textbf{Latency} \\
\midrule
SASRec            & 817.02K & 6.21M & 1.686ms \\
RoTE-SASRec     & 813.76K & 6.32M & 2.423ms \\
\midrule
RPG               & 13.58M & 994.94M & 6.224ms \\
RoTE-RPG         & 13.55M & 994.97M & 8.154ms \\
\bottomrule
\end{tabular}
\end{table}

Table \ref{tab:efficiency} compares the efficiency of backbone models and their RoTE-enhanced variants. RoTE introduces negligible computational overhead, with only marginal increases in FLOPs and inference latency, while maintaining millisecond-level real-time performance. Notably, the slight reduction in parameter count stems from replacing the original positional embedding table with a lightweight computation-based temporal encoding that does not rely on large embedding lookups. Overall, RoTE enables Transformer-based sequential recommendation models to better capture heterogeneous temporal distances among user interactions without incurring significant additional computational cost.

\section{Conclusion}

In this work, we propose RoTE, a plug-and-play rotary time embedding module for Transformer-based sequential recommendation. RoTE models time in a hierarchical and semantically structured manner, enabling models to better capture heterogeneous temporal distances among user interactions. Extensive experiments demonstrate that RoTE consistently improves performance across both traditional and generative backbones while introducing negligible computational overhead. These results indicate that structured temporal modeling is a practical and effective direction for enhancing sequential recommender systems.

\section*{Acknowledgements}

This work was supported by the STI 2030-Major Projects under Grant 2021ZD0201404  and Shenzhen Key Laboratory of New Generation Interactive Media Technology Innovation (ZDSYS2021062\\3092001004).

% \clearpage

\bibliographystyle{ACM-Reference-Format}
\bibliography{SIGIR2026/references}

@String{Computer = "{IEEE} Computer" }

@String{Springer = "Springer-Verlag" }

@inproceedings{sasrec,
  title={Self-attentive sequential recommendation},
  author={Kang, Wang-Cheng and McAuley, Julian},
  booktitle={2018 IEEE international conference on data mining (ICDM)},
  pages={197--206},
  year={2018},
  organization={IEEE}
}

@article{gru4rec,
  title={Session-based recommendations with recurrent neural networks},
  author={Hidasi, Bal{\'a}zs and Karatzoglou, Alexandros and Baltrunas, Linas and Tikk, Domonkos},
  journal={arXiv preprint arXiv:1511.06939},
  year={2015}
}

@inproceedings{caser,
  title={Personalized top-n sequential recommendation via convolutional sequence embedding},
  author={Tang, Jiaxi and Wang, Ke},
  booktitle={Proceedings of the eleventh ACM international conference on web search and data mining},
  pages={565--573},
  year={2018}
}

@inproceedings{bert4rec,
  title={BERT4Rec: Sequential recommendation with bidirectional encoder representations from transformer},
  author={Sun, Fei and Liu, Jun and Wu, Jian and Pei, Changhua and Lin, Xiao and Ou, Wenwu and Jiang, Peng},
  booktitle={Proceedings of the 28th ACM international conference on information and knowledge management},
  pages={1441--1450},
  year={2019}
}

@article{seqrec_study,
  title={Sequential recommendation: A study on transformers, nearest neighbors and sampled metrics},
  author={Latifi, Sara and Jannach, Dietmar and Ferraro, Andr{\'e}s},
  journal={Information Sciences},
  volume={609},
  pages={660--678},
  year={2022},
  publisher={Elsevier}
}

@article{xiao2025unger,
  title={Unger: Generative recommendation with a unified code via semantic and collaborative integration},
  author={Xiao, Longtao and Wang, Haozhao and Wang, Cheng and Ji, Linfei and Wang, Yifan and Zhu, Jieming and Dong, Zhenhua and Zhang, Rui and Li, Ruixuan},
  journal={ACM Transactions on Information Systems},
  volume={44},
  number={2},
  pages={1--31},
  year={2025},
  publisher={ACM New York, NY}
}

@article{tiger,
  title={Recommender systems with generative retrieval},
  author={Rajput, Shashank and Mehta, Nikhil and Singh, Anima and Hulikal Keshavan, Raghunandan and Vu, Trung and Heldt, Lukasz and Hong, Lichan and Tay, Yi and Tran, Vinh and Samost, Jonah and others},
  journal={Advances in Neural Information Processing Systems},
  volume={36},
  pages={10299--10315},
  year={2023}
}

@article{hstu,
  title={Actions speak louder than words: Trillion-parameter sequential transducers for generative recommendations},
  author={Zhai, Jiaqi and Liao, Lucy and Liu, Xing and Wang, Yueming and Li, Rui and Cao, Xuan and Gao, Leon and Gong, Zhaojie and Gu, Fangda and He, Michael and others},
  journal={arXiv preprint arXiv:2402.17152},
  year={2024}
}

@inproceedings{vqrec,
  title={Learning vector-quantized item representation for transferable sequential recommenders},
  author={Hou, Yupeng and He, Zhankui and McAuley, Julian and Zhao, Wayne Xin},
  booktitle={Proceedings of the ACM Web Conference 2023},
  pages={1162--1171},
  year={2023}
}

@inproceedings{rpg,
  title={Generating long semantic ids in parallel for recommendation},
  author={Hou, Yupeng and Li, Jiacheng and Shin, Ashley and Jeon, Jinsung and Santhanam, Abhishek and Shao, Wei and Hassani, Kaveh and Yao, Ning and McAuley, Julian},
  booktitle={Proceedings of the 31st ACM SIGKDD Conference on Knowledge Discovery and Data Mining V. 2},
  pages={956--966},
  year={2025}
}

@inproceedings{amazon,
  title={Image-based recommendations on styles and substitutes},
  author={McAuley, Julian and Targett, Christopher and Shi, Qinfeng and Van Den Hengel, Anton},
  booktitle={Proceedings of the 38th international ACM SIGIR conference on research and development in information retrieval},
  pages={43--52},
  year={2015}
}

@article{rope,
  title={Roformer: Enhanced transformer with rotary position embedding},
  author={Su, Jianlin and Ahmed, Murtadha and Lu, Yu and Pan, Shengfeng and Bo, Wen and Liu, Yunfeng},
  journal={Neurocomputing},
  volume={568},
  pages={127063},
  year={2024},
  publisher={Elsevier}
}

@article{sasrec_revis,
  title={Revisiting Self-Attentive Sequential Recommendation},
  author={Huang, Zan},
  journal={CoRR},
  volume={abs/2504.09596},
  url={https://arxiv.org/abs/2504.09596},
  eprinttype={arXiv},
  eprint={2504.09596},
  year={2025}
}

@article{hou2025survey,
  title={A survey on generative recommendation: Data, model, and tasks},
  author={Hou, Min and Wu, Le and Liao, Yuxin and Yang, Yonghui and Zhang, Zhen and Zheng, Changlong and Wu, Han and Hong, Richang},
  journal={arXiv preprint arXiv:2510.27157},
  year={2025}
}

@inproceedings{p5,
  title={How to index item ids for recommendation foundation models},
  author={Hua, Wenyue and Xu, Shuyuan and Ge, Yingqiang and Zhang, Yongfeng},
  booktitle={Proceedings of the Annual International ACM SIGIR Conference on Research and Development in Information Retrieval in the Asia Pacific Region},
  pages={195--204},
  year={2023}
}

@article{gptrec,
  title={Generative sequential recommendation with gptrec},
  author={Petrov, Aleksandr V and Macdonald, Craig},
  journal={arXiv preprint arXiv:2306.11114},
  year={2023}
}

@article{lopez2024positional,
  title={Positional encoding is not the same as context: A study on positional encoding for Sequential recommendation},
  author={Lopez-Avila, Alejo and Du, Jinhua and Shimary, Abbas and Li, Ze},
  journal={arXiv preprint arXiv:2405.10436},
  year={2024}
}

@inproceedings{covington2016deep,
  title={Deep neural networks for youtube recommendations},
  author={Covington, Paul and Adams, Jay and Sargin, Emre},
  booktitle={Proceedings of the 10th ACM conference on recommender systems},
  pages={191--198},
  year={2016}
}

@inproceedings{fpmc,
  title={Factorizing personalized markov chains for next-basket recommendation},
  author={Rendle, Steffen and Freudenthaler, Christoph and Schmidt-Thieme, Lars},
  booktitle={Proceedings of the 19th international conference on World wide web},
  pages={811--820},
  year={2010}
}

@article{shah2025dsrs,
  title={DSRS: DELIGHT sequential recommender system},
  author={Shah, Syed Tauhid Ullah and Khan, Fazlullah and Yamani, Shirin and Alturki, Ryan and Gazzawe, Foziah and Razzak, Muhammad Imran},
  journal={Engineering Applications of Artificial Intelligence},
  volume={142},
  pages={109936},
  year={2025},
  publisher={Elsevier}
}

@inproceedings{he2016fusing,
  title={Fusing similarity models with markov chains for sparse sequential recommendation},
  author={He, Ruining and McAuley, Julian},
  booktitle={2016 IEEE 16th international conference on data mining (ICDM)},
  pages={191--200},
  year={2016},
  organization={IEEE}
}

@inproceedings{elecrec,
  title={ELECRec: Training sequential recommenders as discriminators},
  author={Chen, Yongjun and Li, Jia and Xiong, Caiming},
  booktitle={Proceedings of the 45th international ACM SIGIR conference on research and development in information retrieval},
  pages={2550--2554},
  year={2022}
}

@article{pan2026survey,
  title={A survey on sequential recommendation},
  author={Pan, Li-Wei and Pan, Wei-Ke and Wei, Mei-Yan and Yin, Hong-Zhi and Ming, Zhong},
  journal={Frontiers of Computer Science},
  volume={20},
  number={3},
  pages={2003606},
  year={2026},
  publisher={Springer}
}

@article{wei2025sequential,
  title={Sequential recommendation system based on deep learning: A survey},
  author={Wei, Peiyang and Shu, Hongping and Gan, Jianhong and Deng, Xun and Liu, Yi and Sun, Wenying and Chen, Tinghui and Hu, Can and Hu, Zhenzhen and Deng, Yonghong and others},
  journal={Electronics},
  volume={14},
  number={11},
  pages={2134},
  year={2025},
  publisher={MDPI}
}

@article{poppel1997hierarchical,
  title={A hierarchical model of temporal perception},
  author={P{\"o}ppel, Ernst},
  journal={Trends in cognitive sciences},
  volume={1},
  number={2},
  pages={56--61},
  year={1997},
  publisher={Elsevier}
}

@inproceedings{www25-gen-rec-tutorial,
  author = {Hou, Yupeng and Zhang, An and Sheng, Leheng and Yang, Zhengyi and Wang, Xiang and Chua, Tat-Seng and McAuley, Julian},
  title = {Generative Recommendation Models: Progress and Directions},
  year = {2025},
  booktitle = {Companion Proceedings of the ACM Web Conference 2025},
}

@article{nasir2023survey,
  title={A survey and taxonomy of sequential recommender systems for e-commerce product recommendation},
  author={Nasir, Mahreen and Ezeife, Christie I},
  journal={SN Computer Science},
  volume={4},
  number={6},
  pages={708},
  year={2023},
  publisher={Springer}
}

@article{hauser2018unix,
  title={UNIX Time, UTC, and datetime: Jussivity, prolepsis, and incorrigibility in modern timekeeping},
  author={Hauser, Elliott},
  journal={Proceedings of the Association for Information Science and Technology},
  volume={55},
  number={1},
  pages={161--170},
  year={2018},
  publisher={Wiley Online Library}
}


\end{document}